# The Effectiveness of a Training Program Based on Health Education to Improve Health Empowerment Level among Refugees in Jordan

Ahmed Al-Sharifin, Department of psychology counselling, College of Education, Yarmouk University, Jordan, ahmed.sh@yu.edu.jo, https://orcid.org/0000-0003-0617-5998

Mu'ayyad Megdadi, Department of psychology counselling, Faculty of Education, Yarmouk University, Jordan, muayyad@yu.edu.jo ORCID: 0000-0003-0767-6153

Amani Shatnawi, Department of Information technology, Faculty of Information Technology and Computer Science, Yarmouk University, Jordan

ashatnawi@yu.edu.jo, Orcid: 0000-0002-5096-9405

Anas AlSobeh, Department of Information Systems, Faculty of Information Technology and Computer Science, Yarmouk University, Jordan, anas.alsobeh@yu.edu.jo , and Information Technology, School of Computing, Southern Illinois University Carbondale, USA. anas.alsobeh@siu.edu,

Orcid: 0000-0002-1506-7924

Aya Akkawi, Department of English Literature, Faculty of Art, Yarmouk University, Jordan

aya-aka @yu.edu.jo

**Abstract**

**Here's the translation of the provided Arabic text into English:**

---

**Abstract**

**Objectives:** The study aimed to evaluate the effectiveness of a health education-based training program in enhancing the level of health empowerment among refugees in Jordan. Health empowerment is a key component to promote health as it enables individuals to control and manage their health outcomes and improve them. Refugees are a vulnerable population group with limited access to healthcare.

**Methodology:** The study sample consisted of 38 refugees in Irbid governorate, Jordan, who were conveniently selected in coordination with some organizations working in the field of asylum in the governorate. They were randomly divided into two groups: an experimental group (n = 19) that received the health education training program, and a control group (n = 19) that did not receive any health education training. The Health Empowerment Scale (HES), a validated tool, was used to collect data from both groups in the pre and post-tests, and a follow-up test was conducted for members of the experimental group only.

**Results:** The results showed a statistically significant increase in the health empowerment scores for the experimental group that received the training program compared to the control group. The mean of the pre-test for the experimental group was (1.97 ± 0.27), and for the control group, it was (1.84 ± 0.21). The post-test mean for the experimental group became (3.88 ± 0.13), while for the control group, it was (1.85 ± 0.20). The follow-up test also demonstrated that the enhanced health empowerment levels were maintained in the experimental group, with no significant difference between the post-test and follow-up scores, indicating the effectiveness of the health education training program in enhancing health empowerment for refugees in Jordan.

**Conclusion:** There are significant responsibilities on healthcare providers and policymakers in Jordan and other countries hosting large numbers of refugees. It's essential to give adequate priority to health education and awareness programs to enhance health empowerment for refugees and improve their health literacy, reflecting on their overall health.

## Introduction

The number of refugees and displaced persons has increased rapidly in recent decades, which has become a major global challenge (Reynolds, 2019). The most important challenges facing host countries have been the provision of protection and basic service such as shelter, food, health care and education (Chemali et al., 2018;

Thoresen et al., 2016). During the past two to three decades, there has been a growing awareness of the psychological, social and material needs of refugees (Esses et al., 2017).

The situation of Syrian refugees in Jordan is similar to refugees in other parts of the world in terms of economic health, social, political, economic, legal and psychological levels. Refugees endure poverty, discrimination, exploitation, lack of access to health services, and encounter social exclusion, in addition to a number of health and psychological problems as a result of their ordeal since the beginning of the crisis ( Al-Qdah & Lacroix, 2016).

No matter how different and varied the circumstances are, refugees remain worthy of respect and deserve the rights that guarantee them a decent life. This was stressed by the United Nations Convention – UNC- of 1951 AD and its Protocol of 1967 AD concerning international human rights (UNHCR, 2018). Jordan is not a signatory of the 1951 UNC or its 1967 Protocol (Francis, 2015). However, Jordan is committed to the Arab Charter on Human Rights of 2004 (League of Arab States, 2004).

Jordan makes its government services SUCH AS education and health Accessible to Syrian refugees (Al shobaki, 2018). Jordan also cooperates with the UNHCR to assist refugees - including Syrians - under the 1998 Memorandum of Understanding (UNHCR, 2018), which states that refugees can enter Jordan without a visa or residence permit and stay. This gives them an opportunity to receive various mental and physical health services by public health care centers.

However, refugees are still experiencing many problems that require periodic evaluation during their asylum in the Jordan. It would provide stakeholders with a sound understanding on which to base their assistance to refugees and find sustainable solutions to their problems. The psychological distress of the refugees was worsened by several factors including economic, social, health and psychological. The most prominent needs are housing, security, food, lack of income, and poor distribution of foreign aid. This lead a number of them to sell food aid to obtain money to fulfill other needs such as accommodation rental. It is noteworthy that these needs fall under economic needs (Wells et al., 2016).

The estimation of the precise number of Syrian refugees in Jordan is not easy because of the registration process and the documents available, however around five million Syrian persons are refugees in Jordan, Lebanon and Turkey, and most of them suffer from economic and health problems (Krafft et al., 2018). Although the international agencies cover the cost of the health services provided to Syrian refugees in

Jordan, though the majority of Syrian refugees live outside the camps, so it couldn't be easy to determine the beneficiaries number (Verme et al., 2015). Jordan is the second host of the Syrian refugees in the region, as it hosts around (10%) of its population, most of them are female; this could be attributed to the nature of females in Arab countries, so they prefer to go to neighboring countries that take the burden of travelling to European countries (Ajlan, 2021; Stevens, 2016).

Syrian refugees in Jordan suffer from different types of physical problems, as some of them are related to war consequences, while others not. About (16%) of Syrian refugees in Jordan have some kind of chronic problems, and it has been indicated that they have some challenges related to personal care behaviors and walking ability, which is considered twice as high as the general population in Jordan. Most of the refugee's medical cases are in need for regular follow ups. More than sixty percent of Syrian refugees in Jordan ask for medical help during their surroundings, while the rest have enough money to travel farther looking for better medical services (Tiltnes et al., 2019).

The integration of refugees into the Jordanian society is not only achieved through environmental factors (e.g., community tolerance and non-discrimination policies), but also through individual factors (e.g., personal skills and abilities) and psychological empowerment factors. Therefore, empowering refugees is critical to their integration into society (The European Council on Refugees & Exiles, 2002). It is the primary objective behind their resettlement as active persons in society (Tomlinson & Egan, 2002).

Although empowering refugees is complex (Steimel, 2017), it is an active and multidimensional process, which is defined as the idea of people having the ability to understand and control themselves and their environments (including health, social, economic and political factors), expand their capabilities and horizons, and elevate themselves to higher levels of achievement and satisfaction (Lee, 2005). This gives them a chance to become agents of change within their communities. It is more than just the acquisition of knowledge and skills by refugees. This involves the acquisition of knowledge and skills that are relevant and important to their lives, and the ability to apply them to real-world situations, communities and environments with the aim of strengthening and improving them (Acree, 2018). Moreover, it contributes to enhancing the ability of the individuals or groups to make meaningful choices and transform those choices into desirable actions and results (Alsop et al., 2006) through the individuals' realization that they possess the knowledge, ability, and competency to be effective members in their societies (Adams, 2003). It is a collective action, in which members of marginalized or oppressed communities unite to create social change (Bookman &

Morgen, 1988). The empowerment actively involves people in making decisions about their well-being, potential, life satisfaction, and outcomes of achieving control over their lives to the fullest extent possible (Nachshen, 2005).

Cowger & Snively (2002) pointed out that individuals are able to make their own choices and decisions, and this does not mean that individuals have the strengths and capabilities to solve their difficult life situations, but that they also increase their strength and contribute to their well-being community. The role of the psychologist and social worker here is to nurture, encourage, help, empower, support, motivate and unleash the power within people; and promote equity and justice at all social levels (Alsop et al., 2006). To do this, the psychologist helps clients clarify the nature of their situations, define what they want, and help them explore alternatives to achieve those desires.

Chamberlain (1997) sees empowerment as more from the perspective of mental health rehabilitation, referring to empowerment as a process that has a number of qualities such as having decision-making power and accessing information and resources (Perkins & Zimmerman, 1995; Zimmerman, 1995, 2000). In contrast, disempowerment includes stress, frustration, and despair (Koegel et al., 2003). From the perspective of humanitarian work with refugees, empowerment is achieved by supporting individuals to meet their needs and achieve their personal goals (Hiegemann, 2013).

Various governmental and non-governmental institutions, as well as international organizations, are making great efforts toward promoting health empowerment among refugees through providing appropriate health education and awareness programs and establishing health centers that provide appropriate health services (Almoshmosh, 2016). The significant increase in the number of refugees, especially between (2014-2015) urged the international institutions to give priority attention to the need for refugees and displaced persons to obtain adequate and appropriate health care services, and this requires concerted efforts at regional and international levels (Samara et al., 2022).

It can clearly be seen that modern life has changed our lifestyle. On the one hand, we have become physically inactive as most of the work can be accomplished without much physical effort. On the other hand, we have become less interested in maintaining healthy practices as a result of living in a state of ease and great comfort, which has led to the emergence of many health problems, especially obesity, and its associated consequences. As a result of modern life, obesity has been viewed as a pandemic that has spread in many societies (Sudikno et al., 2021). The lack to maintain a balance between the desire not to exert much effort and the need to preserve good health exposes individuals to various health issues. Therefore, efforts should be made to introduce positive changes to people's lifestyles with a focus on raising health awareness in the community (Razak et al., 2020).

Practicing a healthy lifestyle is not limited to physical activity, but rather includes any behavior that an individual engages in and affects his/her health such as smoking, inappropriate diet and eating habits, lack of physical activity, alcohol intake, and other behaviors that exert a strong influence on an individual's overall health in a way or another (Leyland & Groenewegen, 2020). Societies place a great deal of importance on behaviors that greatly affect the health of individuals. Norman & Conner (1996) stated that the importance of healthy lifestyle behaviors stems from the fact that statistics indicate that several different unhealthy behaviors have been associated with a marked increase in deaths across the world. In addition, health-related behaviors can be altered, modified, and improved which contributed to reducing related health risks.

Interest in healthy behavior has had several definitions during previous periods; it was called "health education". In the seventies of the last century it was known health education as the process that aims at "bringing about positive behavioral changes in individuals which are conducive to their present and future health" (Glanz et al., 2008, p: 49). In the 1980s, it was defined by (Green et al., 1980) as a set of learning experiences designed to facilitate behavioral changes that improve an individual's overall health. According to the World Health Organization (WHO), as mentioned in (Nutbeam, 1998), " Health education comprises consciously constructed opportunities for learning that are built to contribute to the improvement of individual and community health; this involves some form of communication with different parties such as individuals, schools, and workplaces to improve health literacy, including improving knowledge and developing life skills which are conducive to individual and community health. Health education is not only concerned with the communication of information, but also with fostering motivation, skills, and confidence necessary to take action to improve health, all of which contribute to improving and maintaining health awareness (Nutbeam, 1998).

The term " health literacy" has been used by (Zumbo et al., 2006). It refers to the degree to which people can access, understand, and communicate information to engage with the demands of different health contexts to increase opportunities for promoting and maintaining good health across the life course (Sørensen, 2012). Moreover, promoting health is the process of enabling people to increase control, and improve their health. The focus is moving on individual behavior toward a wide range of social and environmental interventions (WHO, 2016; Shearer & Fleury, 2006). In sum, health education, whether for individuals or societies, is a crucial component in the process of public health management and protection, and despite the different terms being used, they are all identical in meaning (Glanz et al., 2008).

Healthy behavior contributes to public health in all societies, but when it comes to the health of refugees it becomes more complex; this is due to the different life priorities and needs of a refugee. For example, securing the basic needs required for maintaining a decent standard of living for the family is more important than concerns about the various health needs of an individual or even his/her family members. Refugee and asylum issues in modern times are often associated with political issues which reduce the opportunities for growth and development of the refugee community, including health care services for refugees. In addition, refugees continue to face health problems even after integration into the host community; they often find themselves obliged to do jobs lacking proper safety and health protection, or that the members of the host community are usually unwilling to do (Bempong et al., 2019).

Rechel et al. (2011) noted that there has been a lack of understanding of the health needs of refugees and immigrants in the European Union and limitations to the accessibility and delivery of proper healthcare. Moreover, communication between refugees and migrants and their health care providers has been assessed and findings indicated that there are significant limitations to the degree of effective communication and that local healthcare systems failed to adapt adequately to the specific needs of this population. Scholz (2016) pointed out that most of the refugees and migrants originate from countries affected by weak health systems and the various factors of poverty and conflicts in those countries have affected the ability of these systems to provide proper healthcare services.

The World Health Organization (WHO, 2016) believes that the health problems of refugees and migrants are similar to those of the rest of the population, although some groups may have a higher prevalence. The most frequent health problems of refugees and migrants include accidental injuries, hypothermia, burns, gastrointestinal illnesses, cardiovascular diseases, pregnancy- and delivery-related complications, diabetes, and high blood pressure. The suffering of the refugees is also associated with poor living conditions, such as unemployment and poor housing conditions which increase their risks for various health problems and limit their ability to access health care services and meet their basic health needs (Daynes, 2016; Androutsou and Metaxas, 2019).

Planning public health programs that aim at protecting and improving the health of people and their communities requires effective planning and management to assist the program in achieving its overall objectives and to ensure the best chance of success when implementing various public health programs. Health program planners in a targeted population can overcome risks that could present threats to program success by using some health behavior models and theories to understand and explain health behavior. Although this may seem difficult in light of the multiple unhealthy behaviors that should be effectively addressed, these models can help in planning different health programs (Simpson, 2015).

Some studies (Bolzman, 2014; Rizkalla et al., 2020) have indicated that children and older refugees remain at greater risk and even the most in need of health empowerment. Social networks play a significant role in the health care and health empowerment of the elderly. Shearer and Fleury (2006) viewed that social connections have powerful effects on old people's health and wellbeing and can enhance health empowerment and help them deal with their health needs. On the other hand, refugee children, notably, suffer from many health problems such as malnutrition and vitamin deficiencies, especially (Vitamins D, B12, A), in addition to iron deficiency (Shah et al., 2020).

A brief review of the literature reveals that researchers found some studies addressing the aspect of health in refugees, and they could be divided into two divisions:

**First: Studies that are related to healthcare services provided and its obstacles:**

Rousan and colleagues' study (Rousan et al., 2018) sought to determine the how Syrian refugees consider healthcare services provided. The research findings showed that health problems (chronic diseases and mental health) are the main problems facing Syrian refugees in Jordan, while economic problems (from poverty and the inability to secure income) are the real obstacle to accessing health care. It also indicated that refugees perceive increased livelihood opportunities as a potential solution to such problems.

Frost and colleagues' study (Frost et al., 2018) aimed to evaluate the feasibility, acceptability, and perceived impact of a yearlong health education intervention to empower Burmese refugee women living in Houston, Texas in the United States. The evaluation was a formative qualitative study including interviews with Burmese refugee women who participated in the intervention. The final sample of the study consisted of (11) refugee women. After a qualitative content analysis, the results of the study indicated that the motivation to participate in the intervention was impacted by the women's perceived relevance of health education material to Burmese cultural values.

Khan-Gökkaya & Mösko (2020) conducted a study that sought to identify the challenges and barriers faced by refugee health professionals (RHPs) in Germany on their path to reentering their original occupations and to explore the policies used to address those challenges. The study adopted the qualitative research method, and interviews were conducted using 24 participants. The interviews were analyzed using qualitative content analysis. Several main barriers have been identified, most notably lack of language competencies, unfamiliarity with the health care system, skills in dealing with staff members, relationships toward patients, and the experience of discrimination. The results also indicated the need to create an appropriate professional environment to address these barriers. The results indicated that the most prominent personal strategies used by health professionals are to be patient, believe in his/her qualifications at work, and share experiences with staff members. Regarding barriers to dealing with the work team, a positive communication strategy and mutual support is an effective strategy.

Price et al. (2020) showed that the imminent need for greater accessibility to comprehensive sexual and reproductive healthcare services among Syrian refuges residing in Jordan. They used ethnographic case study that consist of (21) male and female Syrian refugees residing in Amman. Participants were selected purposefully. The study interviews examined the experiences of participants regarding their pregnancy outcomes, influencing factors for their reproductive decisions, and gaps in sexual and reproductive healthcare in Amman. The findings indicated great need specifically related to their education on contraceptive methods.

The study of Riza et al. (2020) aimed at identifying the most promising best health practices for migrants and refugees. It was done through performing a scoping review of the international academic literature on effective community-based healthcare models and interventions for migrants and refugees and academic publications of international organizations. To collect data, a systematic search in several databases was conducted, such as EMBASE, PubMed, and Scopus databases, following the Preferred Reporting Items for Systematic Reviews and Meta-Analyses (PRISMA) methodology. Data were extracted from eligible publications included information on general study characteristics, a brief description of the intervention/model, and reported outcomes in terms of effectiveness and challenges. Subsequently, specific criteria were developed to assess these interventions. A total of (118) academic publications were critically reviewed and categorized in the thematic areas to which they belong. The results indicated that the elements of good communication, linguistic barriers, and cultural differences played crucial roles in the effective application of the interventions. In addition, the close collaboration of the various stakeholders, the local communities, the migrant/refugee communities, and the partnerships were key elements in the successful implementation of primary healthcare provision.

Kiani's study (Kiani et al., 2021) aims to illustrate the status of refugees' access to healthcare and the main initiatives to improve their health status in Iran. The authors followed a mixed-method study with two consecutive phases: qualitative and quantitative. In the qualitative phase, the right of refugees to access healthcare services in the Iranian health system was examined through a review of seven documents and interviews with purposively-selected healthcare providers. In the quantitative phase, data on refugees' insurance coverage and their utilization from community-based rehabilitation (CBR) projects were collected and analyzed. Quantitative data were collected from refugees covered under the Universal Health Insurance Scheme and several refugees with physical disabilities who were included in the Community Based Rehabilitation project (CBR). The results indicated that there were international and upstream policies, laws, and practical projects that support refugees' health in Iran. Refugees and immigrants have free access to most healthcare services provided in the health care network in Iran. They can also access curative and rehabilitation services, the costs of which depend on their health insurance status.

**Second: Studies that are related to refugee's health problems:**

Strong et al., (2015) conducted a study aimed at identifying the physical and emotional conditions, dietary habits, coping practices, and living conditions of elderly Syrian refugees in Lebanon. The study was applied to a sample of (210) elderly people aged 60 years and older who were selected from a list of (1800) elderly people receiving assistance and support from The Caritas Lebanon Migrant Center (CLMC) and the Palestinian Women's Humanitarian Organization (PALWHO). Social workers in (CLMC) and (PALWHO) collected qualitative and quantitative information in (2013). The Katz Index of Independence in Activities of Daily Living (Katz ADL) was used. The findings showed a high level of health problems among the study sample. Two-thirds of the older refugees described their health status as poor or very poor. Most reported at least one non-communicable disease, with 60% having hypertension, 47% reporting diabetes, and 30% indicating some form of heart disease. Difficulties in affording medicines were reported by 87%. The results also revealed that diet was inadequate for most older refugees.

Nelson-Peterman et al. (2015) conducted a study aimed at identifying the factors influencing public health of a sample of Cambodian refugee women in the United States. To achieve the objectives of the study, the researchers designed a questionnaire that involves a set of factors affecting health, and it was applied to a sample of (160) women in the city of Massachusetts, aged (35 to 60) years. The finding showed that smoking rates were low (4%), and walking exercise rates were also low. 50% of the participants indicated that they do not practice walking two times a week. In addition, (73%) of them suffer from obesity according to the criteria of the World Health Organization.

Khan and Amatya's study (2017) also aimed to provide an overview of the health conditions of refugees and potential challenges from the rehabilitation perspective. To achieve the objectives of the study, the authors conducted the study using medical and health science electronic databases and internet search engines (2001-2016). Both authors independently selected studies. Due to the heterogeneity of the selected studies, a narrative analysis was performed for best-evidence synthesis to outline the current health and rehabilitation status of refugees. The results indicated that infectious diseases that require treatment in refugees are a minority, while some non-communicable diseases, such as musculoskeletal conditions, are more prevalent. The results also indicated that one of every six refugees suffers from a physical health problem severely affecting their lives, and nearly two-thirds experience mental health problems. The study also emphasized the important role of health rehabilitation for refugees.

Rizkalla et al., (2020) also conducted a study that aimed to examine Syrian refugee mothers' accounts of the physical and mental health of their children being affected by war traumas and displacement challenges. The study was conducted by recording interviews using 23 mothers aged (21 to 55) residing in Jordan after obtaining consent from participants. Using a narrative approach in the data collection and analysis, some major themes were identified related to the diverse war-related traumatic events the children experienced while they were in Syria, or during their escape journey, and the difficulties the children are currently facing. The study provides empirical findings that war events and displacement have adversely impacted both the physical and mental health of Syrian refugee children and that many of them **suffer to secure** a decent life. Children experienced sexual violence. Mothers also narrated that displacement exposed children to poverty, decreased quality of food, hostility from local peers, and some educational and recreational challenges.

Previous studies clearly indicate that health problems are relatively high among refugees (Nelson-Peterman et al., 2015; Strong et al., 2015). Some of these studies tried to evaluate health education interventions (Frost et al. al., 2018), while some other studies aimed at investigating the refugees' access to health services (Doocy et al., 2016; Kiani et al., 2021). The studies also involved different categories of the refugee community; some studies included refugee women (Nelson-Peterman et al., 2015; Frost et al., 2018), others included elderly refugees (Strong et al., 2015), while some tried to investigate the impact of displacement on the general health of children (Rizkalla et al., 2020). It is clear from the above review that previous investigations have focused on multiple health issues of refugees which is very important in understanding the refugees' health conditions.

However, previous research did not directly investigate the impact of a health education program on the target refugee community, and this enhances the significance of the current research.

**Significance of the Study**

This study comes to address one of the major issues in the lives of refugees, which is the health condition of Syrian refugees. Providing health education to Syrian refugees contributes to increasing the awareness of health issues for all population categories (children, adults, the elderly, women), which helps them seek medical assistance from specialized medical resources. It is also important to consider that physiological needs are the most basic needs at the lowest level of the individual hierarchy of needs which should be fulfilled before moving on to other, more advanced needs. Rearranging priorities based on the hierarchy of needs helps to restore the state of balance among refugees.

Although countries and various international organizations are committed to providing refugees with health support and medical assistance needed, the results do not meet the expectations. This study aims to explore the Syrian refugees' health status and healthcare in Jordan, which will provide insights for decision-makers and international organizations into the development of health empowerment programs for Syrian refugees in Jordan. Achieving successful results in this study will contribute to the possibility of generalizing the findings to refugee communities outside Jordan and in other countries of the world.

**Statement of the Problem**

The problem being addressed in this study emerges from the findings of some studies that reported the prevalence of various health problems in refugees and migrants (Nelson-Peterman et al., 2015; Strong et al., 2015; Khan & Amatya, 2017). Pinheiro and Jaff (2018) indicated that Syrian refugees in Jordan face a range of difficulties accessing health services, and this may clearly reflect their need for health education and promotion programs aimed to advance knowledge of public health and public health resources. Sharp et al., (2018) indicated a high prevalence of chronic diseases among Syrian refugees in Jordan than in their Jordanian peers. Rechel et al. (2011) pointed out that more attention and focus should be given to the health condition of refugees to get adequate knowledge and a better understanding of their specific health needs. In addition, Khan and Amatya's study (2017) pointed to the effectiveness of rehabilitation interventions in improving health-related quality of life and community health.

Although Jordan has an advanced healthcare system that is one of the best in the region, the large number of refugees in Jordan inevitably impacts the efficiency of the healthcare system and its ability to provide adequate health services (Pinheiro & Jaff, 2018). Health issues continue to be a vexing problem in all societies. Countries and governments give particular emphasis on providing quality health services to all citizens. However, when it comes to refugees and the various challenges they face, refugees' healthcare becomes more important. Hence, this study aims to present a proposed health education program for Syrian refugees in Jordan.

**Research Questions**

This study aims to answer the following questions:

1. How effective is a training program based on health education to raise the level of health empowerment among refugees in Jordan?
2. How effective is a continuing health education-based training program in raising the level of health empowerment among refugees in Jordan?

**Study Hypotheses**

1. The first hypothesis: There are no statistically significant differences at the level of significance ($\alpha = 0.05$) between the average scores of the members of the experimental group and the average scores of the members of the control group on the health empowerment scale.
2. The second hypothesis: There are no statistically significant differences at the level of significance ($\alpha = 0.05$) between the average performance of the experimental group in the post-test of the health empowerment scale (HES) and their average scores in the follow-up test a month after the end of the program.

**Limitations of the study**

The possibility of generalizing the study findings is determined according to multiple conditions: the demographic characteristics of the study sample of refugees in Irbid Governorate - Jordan, who were found to have poor health behaviors. The possibility of generalizing the study findings is also determined according to implementing a quasi-experimental research design with two groups (i.e., the experimental and control groups) with random assignment, and pre-post and follow-up measurements, and according to the psychometric properties of the data gathering tools used in the study, which were based largely on self-report data and participants' results on the health behavior scale. In addition to the elements of the training program used in this study and the Spatio-temporal model used throughout the study procedures.

**Key Terms**

Health Empowerment: Health education is the process that aims at bringing about positive behavioral changes in individuals which are conducive to their present and future health (Glanz et al., 2008). It is defined for the purpose of the current study as the degree respondents obtained on the scale used to measure health empowerment.

Syrian refugees: **The**y are the Syrian individuals who have fled the country over the course of the Syrian Civil War starting in (2011) seeking safety and stability in Jordan to protect themselves and their families from different acts of violence. For the purposes of the current study, they are defined as Syrian refugees participating in the health education- training program.

**Methods and Procedures:**

**Study Sample**

The study sample includes some of refugees in Irbid Governorate, Jordan, during the period from 1/10/2021 to 3/30/2021. Several associations and organizations that provide psychosocial and health support services to refugees in Irbid have been selected using (Convenience Sampling). They have cooperated with the researchers in announcing the training program, which was implemented in one of these associations' halls. Health Empowerment Scale (HES) was adapted and applied to all beneficiaries of the various services. Subjects were randomly selected from large numbers of Syrian refugees who were getting assistance from the associations and organizations reached out by the researchers. A sample size of (n= 38) refugees who had the desire to participate in the program to represent the study sample and who got the lowest score on (HES) was chosen. Subjects were randomly divided into two groups: The experimental group (n = 19) refugees (the group who took the training program) and the control group (n = 19) refugees (without any training intervention).

## Study Tools

**First: Health Empowerment Scale (HES):** Serrani Azcurra DJL (2014). Elders Health Empowerment Scale. Spanish adaptation and psychometric analysis was adapted because of its excellent validity and reliability criteria while reflecting the attributes of health empowerment. The scale consists of (8) items to which the refugee responds to a five-point scale and has psychometric properties (validity and reliability) used in similar studies.

### Psychometric Properties of the Scale in its Current Form

To assess the psychometric properties of (HES) in its current form, the scale has been translated from English into Arabic by a group of specialized faculty members to assess the translation validity. Their comments about the validity, clarity, comprehensibility, and conceptual and operational comparability of the translation were taken into consideration.

### Face Validity

The translated version of the (HES) was presented to (10) reviewers who were specialist professors and were asked to review the validity, language clarity, and conceptual equivalence of the translated items for the study sample. In light of the reviewers' opinions, the proposed modifications were made on the items of the scale in terms of modes of expression and wording. 80% of the reviewers' consensus agreement on drafting the scale items was adopted by researchers and thus the scale became composed of (8) items.

### Indicators of Construct Validity

An initial sample (n= 30) was used to measure the construct validity. This sample of refugees was from outside the target group. Indicators of construct validity were measured using item coefficient correlation with dimension. It was noted that the coefficient correlations values of the items with the total score on the scale ranged between (0.48 - 0.63). The acceptance of item construct validity was demonstrated by the fact that its correlation coefficient should not be less than (0.20), as indicated by Zuckerman and colleagues (Zuckerman et al., 2004). Thus, all scale items were accepted.

### Scale Reliability

Internal Consistency Reliability: Internal consistency of the (HES) was indicated with a Cronbach's alpha coefficient of (0.86) in the initial sample.

**Test-Retest Reliability:** The **reliability of (HES) was** tested by using **test-**retest.

To assess the test-retest reliability, the scale was re-applied to the initial sample (n= 30) two weeks later. It was tested with the Pearson correlation coefficient which was (0.89).

### The Training Program

The training program used in the current study was designed after considering the Syrian refugees' different needs related to their health problems. The program was designed based on the cooperation and different opinions of many specialists and refugees who were interviewed by the researchers. The program was constructed based on similar programs that targeted the study variables along with referring to several scientific resources and references specialized in psychological support and health empowerment.

The program implementation period was carried out for five days over the period of three weeks during which time 10 training sessions were held. Each session was around (120) minutes, depending on the content of each session. Yet, due to Covid-19 restrictions, the program implementation period was reduced to three days with intensive training sessions per day.

The training program is designed based on the (ADDIE) model for instructional design. This model includes five stages characterized by a logical sequence in the design of the program, as indicated by Davis (Davis, 2013), which are:**Analysis**: The authors analyzed the characteristics of the refugees, the nature of the training centers in terms of the physical environment, the training staff and their level of skills and competencies. The nature of the programs offered to refugees within the center and organizations was also analyzed to ensure that there was no overlap in the sessions of the program based on psychosocial support.

**Design**: The articles of the training content were identified, and the general and specific objectives of the program were defined. Studies and research that dealt with support programs for refugees were reviewed. The ways and means of preparing the program were identified, and how to apply it procedurally to the members of the current study sample.

**Development**: At this stage, the training program based on health education was developed as follows:

The program aims to investigate the effectiveness of health education in improving the level of health empowerment among refugees. This is to enhance their ability to deal with the health problems they face in their lives and help them become aware of self-health and its requirements and understand the foundations of reducing communicable and infectious diseases, and the foundations of limiting their complications. The program also aims to develop their skills in general health care for children, pregnant women, and the elderly. Each session includes specific objectives, techniques, exercises, activities, and homework.

**Tools and methods used**: The following tools and methods were used in the program: Dialogue and discussion, reinforcement, role-playing, questions, self-affirmation, feedback, self-monitoring, self-control, positive self-talk, homework, storytelling, board, toys, boosters, crayons, cuddly stimuli, sketchbooks, wooden pens, and boards.

**The content of the program in its initial form**: The program initially consisted of (5) sessions, one session per week, and the duration of each session was (90) minutes.

**Indications of the validity of the program:** To verify the validity of the program, it was presented to specialized 10 reviewers. The reviewers recommended the following: Correcting some language errors, increasing the time period for the sessions, changing some topics of the sessions, modifying the content of some sessions, and increasing the number of sessions.

**The content of the program in its final form**: All reviewers' recommendations were taken into consideration and the necessary changes were made such as correcting the language errors and increasing the time period for one session. Thus, the number of sessions of the program in its final form reached (10) training sessions, each session lasting (120) minutes. Here is a summary of the program:
- The program is based on a basic idea that lies in the strength of the link between efforts to introduce positive changes in the lifestyle of individuals, with a focus on spreading awareness of health behavior in the community, and this enhances the refugee individual's sense of health empowerment as a result of psychosocial support based on health education. It is included in an organized manner a number of sequential goals, giving an integrated perspective, adherence to which leads to optimally achieving health empowerment. The objectives are distributed over the ten sessions of the program. The sessions of the program include a wide variety of sensory and cognitive exercises that seek to achieve the objectives included in these sessions. These objectives are as follows: Knowing the current health conditions of refugees, knowing the importance of protecting the safety of basic sources such as food and water, awareness of self-health and its requirements, and the foundations for its care. The goals also included introducing the foundations of reducing communicable and infectious diseases, limiting the exacerbation of chronic diseases, achieving an understanding of the foundations of public health care for children, women, pregnant women, and the elderly, and introducing appropriate ways to obtain health services.

**Implementation**: At this stage, the program was implemented at Yarmouk University by specialists who went through preparation to master the content of the program. The program sessions, their contents, the dates of their implementation, and the training were carried out in cooperation between researchers and trainers. Note that the trainers were selected based on their tenure and health knowledge. Where the program was applied to the experimental sample of (19) male and female refugees, where the application was carried out in a special training hall, which is prepared in accordance with the conditions of application and the characteristics of the refugees.

**Evaluation**: The evaluation process for the program was carried out continuously at all stages of designing and implementing the program. After each stage of program development, the outputs were presented to a group of specialized reviewers to evaluate the quality of the program design and the suitability of its objectives to the target group. The feedback provided by the trainers who were trained to implement the program was also taken into consideration, in addition to the feedback from the refugees.

**Study variables**

**The independent variable:** It is the variable being tested and measured in the two study groups: The experimental group (which received the training program) and the control group (which did not receive the training program). **The dependent variable:** It is the level of refugees' health empowerment.

**Statistical Analysis**

After collecting data from the study sample, the researchers calculated the means and standard deviations using a pre, post, and follow-up measures, ANCOVA methods, and (T-Test).

**Study Results**

**First: Results related to the first question.**

To answer the first question: How effective is a training program based on health education to raise the level of health empowerment among refugees in Jordan?
the following hypothesis was set: There are no statistically significant differences at the level of significance (α = 0.05) between the average scores of the members of the experimental group and the average scores of the members of the control group on the health empowerment scale. To verify the validity of the study's first hypothesis which states that "there are no statistically significant differences at the level of significance (α = 0.05) between the average scores of the experimental group and the control group in the (HES) post-test", the arithmetic means and standard deviations of the pre and post measures of the level of refugees' health empowerment were calculated according to the group variable (experimental and control), Table (1) illustrates these results.

**(Table 1): The means and standard deviations of the responses of the study groups (experimental and control) (n= 38) on (HES) in the pre and post-tests according to their variables**

| GROUP | | pre | post |
|---|---|---|---|
| EXPERIMENTAL | Mean | 1.9737 | 3.8837 |
| | N | 19 | 19 |
| | Std. Deviation | .27091 | .13817 |
| CONTROL | Mean | 1.8411 | 1.8595 |
| | N | 19 | 19 |
| | Std. Deviation | .21494 | .20748 |
| TOTAL | Mean | 1.9074 | 2.8716 |
| | N | 38 | 38 |
| | Std. Deviation | .25039 | 1.04032 |

Table (1) shows that there are significant differences between the means of the experimental and control groups (n = 30) on (HES). To verify the significant differences, the one-way ANCOVA analysis was used after taking into account (HES) pre-test for both groups as a common variable for the presence of variance in the (HES) post-test. Table (2) illustrates these results.

**(Table 2) One-way analysis of covariance (ANCOVA) of (HES) post-test according to the group variables**

| Source | Type III Sum of Squares | df | Mean Square | F | Sig. | Partial Eta Squared |
|---|---|---|---|---|---|---|
| Corrected Model | 39.292[a] | 2 | 19.646 | 913.956 | .000 | .981 |
| Intercept | 2.533 | 1 | 2.533 | 117.816 | .000 | .771 |
| a_pre | .366 | 1 | .366 | 17.036 | .000 | .327 |
| GROUP | 34.195 | 1 | 34.195 | 1590.820 | .000 | .978 |
| Error | .752 | 35 | .021 | | | |
| Total | 353.391 | 38 | | | | |
| Corrected Total | 40.044 | 37 | | | | |

a. R Squared = .981 (Adjusted R Squared = .980)

Table (2) shows that there are statistically significant differences ($\alpha = 0.05$) between the mean of (HES) pre and post measurement among the refugees participated in the study. To determine which of the two study groups (experimental and control) has the apparent differences, the modified mean of (HES) among refugees was calculated based on the group and its standard errors, as shown in Table (3).

**Table (3): The modified mean and standard errors of (HES) post-test according to the group variable.**

| Variable | Group | Modified arithmetic mean | Standard error |
|---|---|---|---|
| Health Empowerment | **Experimental** | 3.856 | 0.034 |
| | **Control** | 1.887 | 0.034 |

Table (3) shows that the apparent differences are in favor of the experimental group which received training on health empowerment, compared to the control group which did not receive any training. This has led the researchers to reject the null hypothesis in favor of accepting the alternative hypothesis which indicates that there is a difference between the two groups. Since the impact of the program has reached (97.8%), this indicates that there is a significant impact of the Health empowerment training program based on (HES). This also indicates that this training program has a significant impact on the health improvement of the refugees who participated in the training program.

**Second: Results related to the second question**.

To answer the second question: How effective is a continuing health education-based training program in raising the level of health empowerment among refugees in Jordan?, the following hypothesis was set: The second hypothesis: There are no statistically significant differences at the level of significance ($\alpha = 0.05$) between the average performance of the experimental group in the post-test of the health empowerment scale (HES) and their average scores in the follow-up test a month after the end of the program.

To verify the validity of the second hypothesis of the study, which states that "there are no statistically significant differences ($\alpha = 0.05$) between the average performance of the experimental group on (HES) in the post-test, and their average scores on the same scale in the follow-up test one month after the end of the test", the mean and standard deviations were calculated and the Paired-Sample t-test was used to indicate the experimental group's performance in the post and follow-up tests on (HES).

Table (4) shows the means and standard deviations of the post and follow-up tests and the results of the "T-Test" for the differences between the two scales of the post and follow-up tests on (HES) among the experimental group, which were (19) refugees.

**Table (4): The means, standard deviations, and "t-test" results for the differences between the two scores of the post and follow-up tests on (HES) for the members of the experimental group (n = 19)**

| Variable | Measurement | Arithmetic means | Standard deviation | (t) value | Statistical significance |
|---|---|---|---|---|---|
| Health Empowerment | Pre | 3.8837 | .13817 | 1.797 | 0.089 |
|  | Follow-up | 3.8463 | .13188 |  |  |

*Statistical function at the level of significance ($\alpha \geq 0.05$)

Table (4) indicates that there are no statistically significant differences (α = 0.05) between the scores of the members of the pilot study who participated in the training program on (HES). This indicates that the refugees have met the training objectives due to their enrollment in the program.

**The Discussion**

It is possible to explain the previous result through the goals that empowerment seeks to fulfill BY enhancing the capabilities of individuals to make meaningful choices and transform those choices into desired actions and results (Alsop et al., 2006), or through the individual's realization that he or she possesses the knowledge, ability and capacity to be an effective member of their lives and society (Adams, 2003). This is what the educational program sought to achieve through training sessions, which aimed to enhance the refugees' different knowledge about physical diseases and ways to protect them, and to introduce healthy behaviors. Thus, refugees may have acquired the ability to make decisions about their health and disease conditions.

The results obtained can be interpreted based on the foundations of the educational program, whereby empowerment is seen as an active process through which refugees attempt to affect change in relationships, attitudes, or contexts (Brodsky & Cattaneo, 2013). Empowerment means that individuals can act to prevent problems, gain or restore the ability to interact with the social environment, and expand the resources available to meet their needs (Long et al., 2006). It actively involves people in making decisions about their well-being, potential, life satisfaction, and outcomes for achieving maximum control over their lives (Nachshen, 2005), and this is consistent with the various goals that the program aims to achieve.

Moreover, it is possible to explain the result in the light of what Cowger & Snively (2002, p.110) had stated by indicating that promoting empowerment means believing that individuals are able to make their own choices and decisions, and this does not mean that individuals have strengths and capabilities to solve difficult life situations, but they also increase their power and contribute to the welfare of society by doing so. The role of the psychologist and social worker here is to nurture, encourage, help, empower, support, motivate and unleash the power within individuals; to highlight the strengths available to them in their own environments; and to promote equity and justice at all levels in the society (Alsop et al., 2006). To do this, the psychologist helps to clarify the nature of their situations, determine what they want, and help them explore alternatives to achieve those desires. This, in turn, may support refugees unleash their inner strengths?

The result can also be interpreted in the light of what the program emphasized when it comes to the basic pillars included in empowerment through the control of two forces, namely: the internal force, which includes the internal psychological factor and includes the sense of control, efficiency, responsibility, future orientation; and the external force, which includes the social factor covering control of Sources of personal power and work, and the ability to control the social environment (Jha & Nair, 2008).

It is possible to explain the result in light of the knowledge of the characteristics of the experimental group and their strong desire to purchase the educational program. It aims to stimulate the psychological empowerment process through the programs it offers, and the most important of these programs are the psychological and social support programs. Fawcett and others (Fawcett et al., 1994) indicated that according to the contextual model of empowerment, environmental and personal variables can facilitate or hinder empowerment. According to the theoretical literature, psychosocial factors (eg, social support, self-esteem, internal locus of control, self-efficacy, etc.) contribute to empowerment. According to Rousan et al. (2018) health problems (chronic diseases and mental health) are the main problems facing Syrian refugees in Jordan.

The current study sought to examine the effectiveness of a training program based on health education in raising the level of refugees' health empowerment in Jordan. The results of the first hypothesis examination indicated that there were statistically significant differences at the level of significance ($\alpha = 0.05$) between members of the experimental group and the control group; and that the differences were in favor of the experimental group which received training on health empowerment, compared to the control group which did not receive any training. There was a significant improvement in healthy behaviors among the experimental group which received training on health empowerment.

This result emphasizes the findings of the previous descriptive studies, which indicated the refugees' need for health education to practice healthy behaviors in their daily lives. The study of Shearer & Fleury (2006) indicates that social connections have powerful effects on old people's health and wellbeing and can enhance health empowerment and help them deal with their health needs. The results of the current study emphasize the need to develop educational programs for refugees and are consistent with the study findings of Nelson-Peterman et al. (2015) which indicated a decline in healthy behaviors of Cambodian refugees in the United States. The results showed that walking exercise rates were low as (50%) of the participants which indicated that they do not practice walking two times a week. In addition, (73%) of them suffer from obesity according to the criteria of the World Health Organization. The results of the current study are also consistent with the study findings of Strong and colleagues (Strong et al., 2015), which showed a high level of health problems among the study sample members. Two-thirds of older refugees described their health status as poor or very poor. Most reported the existence of at least one non-communicable disease, with 60% having hypertension, 47% reporting diabetes, and 30% indicating some form of heart disease. Difficulties in affording medicines were reported by 87%. The results also revealed that diet was inadequate for older refugees.

The positive impact of the educational program on the experimental group and its effect on improving healthy behaviors can be explained in light of what was indicated by many researchers. Health education is a process that aims at bringing about positive behavioral changes in individuals which are conducive to their present and future health. In addition, Health education comprises consciously constructed opportunities for learning that are built to contribute to the improvement of individual and community health (Nutbeam, 1998) through accessing and understanding information to engage with the demands of different health contexts to increase opportunities for promoting and maintaining good health across the life-course (Zumbo et al., 2006). This is what the current program has provided, which has enhanced opportunities to practice healthy behaviors by enabling individuals to understand their health and the factors affecting it, and thus increase their ability to control various aspects of health.

The positive result of this research project can be explained in light of the flexibility and diversity of the educational topics included in the program, which helped satisfy the needs of the different categories of the sample. It was also possible to employ various methods and different activities which contributed to achieving the desired outcomes, having fun, and breaking the deadlock. This is consistent with Rechel et al., (2011) who indicated that a lack of adequate understanding of refugees' health needs can have a clear negative impact on the public health of refugees.

Given that this research is targeting refugees, the positive results of this research project can be viewed through its contribution to enhancing the refugees' ability to make decisions about health behaviors as a result of the program's implementation by health care professionals which enhanced the communication and consultation with them in various specific health issues because effective communication between health care providers, refugees, and migrants help to improve practicing healthy behaviors and improve public health in general. The health needs and issues of refugees in Jordan vary greatly depending on their subjective culture, and on their confidence to adapt to the living conditions they encountered in Jordan. Therefore, group and open discussions will be a way to solve problems and help them feel that they are not the only ones who suffer from such health problems.

The role of the program providers during the implementation process, their personal characteristics, and their ability to lead the group effectively are all factors that could have contributed to achieving the objectives of the program and achieving positive outcomes. In terms of the program implementation procedures, the procedures were planned in light of the refugees' needs and they were planned before designing the educational program through cooperation with various healthcare centers and institutions providing different health services to refugees. This encouraged the participants to interact with the educational topics included in the program and increased the effectiveness of the implementation process. Our contribution aims to achieving the program's objectives and reaching positive results was the role of the program providers in the implementation procedures, their personal characteristics, and their ability to lead the group; In terms of implementation procedures, the procedures identified in light of the refugees' needs were adhered to, which were counted before designing the educational program through cooperation with various health centers and bodies concerned with providing health services to refugees. This encouraged the participants to interact with the topics presented and increased the effectiveness of the application. The role of the program providers in asking effective questions played an important role in encouraging participants to raise their problems, taking into account privacy and the cultural context because a number of program implementers were doctors specializing in some specialties; and in terms of personal characteristics, the providers of the educational program enjoyed flexibility and acceptance, respect, empathy, unconditional acceptance, and originality. This, in turn, provided a safe environment in the group, which contributed to the participants' commitment to the rules agreed upon in the group since the start of the program, and encouraged them to implement the activities, which worked to enhance confidence among them and achieve the goals and expectations.

It could be said that the general outcomes of this study are consistent with some previous studies (Yang et al., 2015), even though it was conducted on immigrant women in Taiwan and not on Syrian refugees, and the study of Frost and colleagues (Frost et al., 2018), which was conducted on immigrant in Texas state in the USA.

The result related to the study's second hypothesis stated that there are no statistically significant differences at the level of significance ($\alpha = 0.05$) between the average performance of the experimental group in the post-test of the health empowerment scale (HES) and their average scores in the follow-up test a month after the end of the program. This result indicates the stability of the impact of the program on the participants in the experimental group and their ability to maintain therapeutic gains, and it also indicates the impact of the educational program in improving the level of health behaviors even after a month since the completion of the program.

The stability of the program's contribution to improving the level of health behaviors can be explained that the program, with its various components and elements, helped the participants in the experimental group gain many skills during the program sessions, as the program strengthened the participants' ability to think about the consequences of their decisions associated with health behaviors which have a big impact on their health conditions in the future. This helped to raise health awareness and emotionally prepare them for change, thus increasing the participants' awareness of their problem and their desire to solve it.

In addition, the educational program provides several therapeutic alternatives, which allows the participants to make the appropriate decision and enhance their thinking about the consequences of their various decisions. It is also useful in providing realistic solutions and helping participants find the medical care centers that provide all healthcare services to refugees free of charge, and this helps in maintaining therapeutic gains. The participants' awareness of their health problems also contributes to enhancing the awareness of their strengths, and thus the possibility of employing these aspects in maintaining therapeutic gains.

**Study Recommendations**

Based on the results of the study, the following recommendations are made:
- Training those working with refugees on the methods of implementing the program to use it with other participants and recommending its application in the field.
- Adopting the program used in this study to deal with refugees in organizations specialized in refugees issues.
- Directing the attention of international and local organizations to the importance of developing programs to raise the level of health empowerment among refugees.
- Applying such program on other refugees (e.g., males working in the private sector, women researchers/unemployed, informal workers, etc.) and examining the effectiveness of the program in order to generalize its results.

- Holding training workshops for psychological counselors, social, and educational institutions to train them on the use of the health education program, and how can they employ it to deal with various refugees problems.

**Acknowledgment and financing**

We gratefully acknowledge the interest and efforts of the persons and institutions that have participated in this investigation, financed by the European Union through the Horizon 2020 program for the project "Reshaping Attention and Inclusion Strategies for Distinctively vulnerable people among the forcibly displaced (RAISD)" (grant agreement 822688).

**References**

Acree, K. (2018). *The Impact of an Introduction to Community Psychology Course on Refugee Sense of Empowerment*. American University in Cairo, Master Thesis. AUC Knowledge Fountain.

Adams, R. (2003). *Social Work and Empowerment*. Palgrave Macmillan.

Ajlan, A. (2021). Divorce and Domestic Violence Among Syrian Refugees in Germany. *Journal of interpersonal violence*, 0886260520985488.

Al-Shraifin, A., Arabiat, R.B., Shatnawi, A., AlSobeh, A. and Bahr, N., 2023. The effectiveness of a counseling program based on psychosocial support to raise the level of economic empowerment among refugees. *Current Psychology*, pp.1-10.

Al shobaki, Wa'ed (2018). A Synopsis of the Jordanian Governance System in the Management of the Syrian Refugee Crisis. *Journal of Intercultural Studies,* 39 (5), 596–603. doi:10.1080/07256868.2018.1508007

Almoshmosh, N., Mobayed, M., & Aljendi, M. (2016). Mental health and psychosocial needs of Syrian refugees and the role of Syrian non-governmental organisations. *BJPsych International,* 13(4), 81-83.

Al-Qdah, T. A. K.; Lacroix, M. (2016). Syrian refugees in Jordan: Social workers use a Participatory Rapid Appraisal (PRA) methodology for needs assessment, human rights and community development. *International Social Work*, *5* (14), 2- 11. doi:10.1177/0020872816673889

Alsobeh, A. and Aloudat, A., 2022. The Repercussions of the COVID-19 Pandemic on Higher Education and its implications for Syrian Refugees Students (An Analytical Descriptive Study). *Dirasat: Human and Social Sciences*, *49*(5), pp.150-166.

Alsop, R., Bertelsen, M., & Holland, J. (2006). *Empowerment in practice: from analysis to implementation*. The World Bank. Washington, DC.

Androutsou, L., & Metaxas, T. (2019). Health Systems that Meet the Health Needs of Refugees and Migrants. *Journal of Developing Societies*, 35(1), 83-104.

AL-Refai, N.S., Albahr, N.A., Migdady, A.F., AL-Mifleh, E.I., Shatnawi, A. and AlSobeh, A., 2022 فاعلية برنامج تدريبي مستند إلى الدعم النفسي الاجتماعي في رفع مستوى التمكين النفسي لدى اللاجئين. *International Journal of Psychological and Educational Research*, *1* (2).

Bempong, N. Sheath, D., Seybold, J., Flahault, A., Depoux, A., & Saso, L. (2019). Critical reflections, challenges and solutions for migrant and refugee health: 2nd M8 Alliance Expert Meeting. *Public health reviews*, *40*(1), 1-12.

Bolzman, C. (2014). Older refugees. *The Oxford handbook of refugee and forced migration studies*, 409-420.

Bookman, A., & Morgen, S. (1988). *Women and the politics of empowerment*. Philadelphia: Temple University Press.

Brodsky, E., & Cattaneo, B. (2013). A transconceptual model of empowerment and resilience: Divergence, convergence and interactions in kindred community concepts. *American Journal of Community Psychology, 52(1)*, 333–346.

Chamberlain, M. (1997). Power in action: Empowerment of Indigenous communities. A midwifery run birthing centre in the Canadian Arctic. *The Truth, Virtue and Beauty of Midwifery*, 85-93.

Chemali, Z., Borba, C., Johnson, K., Khair, S., & Fricchione, L. (2018). Needs assessment with elder Syrian refugees in Lebanon: implications for services and interventions. *Global Public Health,13*(9), 1216–1228.

Cowger, C. D., & Snively, C. A. (2002). Assessing client strengths: Individual, family, and community empowerment. *The strengths perspective in social work practice*, 106-123.

Davis, A. (2013). Using instructional design principles to develop effective information literacy instruction: The ADDIE Model. *College and research Libraries news, 74* (4), 205-207.

Daynes, L. (2016). The health impacts of the refugee crisis: a medical charity perspective. *Clinical medicine*, 16(5), 437-440

Doocy, S., Lyles, E., Akhu-Zaheya, L., Burton, A., & Weiss, W. (2016). Health service utilization and access to medicines among Syrian refugee children in Jordan. *The International journal of health planning and management*, *31*(1), 97-112.

Esses, M., Hamilton, K., & Gaucher, D. (2017). The global refugee crisis: empirical evidence and policy implications for improving public attitudes and facilitating refugee resettlement: the global refugee crisis. *Soc Issues Policy Rev, 11*(1), 78–123.

Francis, A. 2015, September 21. Jordan's refugee crisis. *Carnegie Endowment for international Peace*. http://carnegieendowment.org/2015/09/21/jordan-s-refugee-crisispub-61338

Frost, E. Markham, C. Springer, A. (2018). *Refugee Health Education: Evaluating a Community-Based Approach to Empowering Refugee Women in Houston.* Texas Elizabeth Leah Frost Christine Markham.

Glanz, K., Rimer, B. K., & Viswanath, K. (Eds.). (2008). *Health behavior and health education: theory, research, and practice*. John Wiley & Sons.

Green, L. W., Kreuter, M., Deeds, S. G., & Partridge, K. B. (1980). Health education planning: A diagnostic approach. In *Health education planning: a diagnostic approach* (pp. 306-306).

Hiegemann, V. (2013). *Empowerment through education: The case of adult refugees in Cairo. The Center for Migration and Refugee Studies.* The American University in Cairo, Cairo, Egypt.

Jha, S. S., & Nair, S. K. (2008). Influence of locus of control, job characteristics and superior-subordinate relationship on psychological empowerment: A study in five star hotels. *Journal of management Research, 8*(3), 147.

Khan, F., & Amatya, B. (2017). Refugee health and rehabilitation: challenges and response. *Journal of rehabilitation medicine*, *49*(5), 378-384.

Khan-Gökkaya, S., & Mösko, M. (2020). Labour market integration of refugee health professionals in Germany: challenges and strategies. *International Migration*, *59*(4), 105-126.

Kiani, M. M., Khanjankhani, K., Takbiri, A., & Takian, A. (2021). Refugees and Sustainable Health Development in Iran. *Archives of Iranian Medicine (AIM)*, *24*(1), 27-34.

Koegel, R. L., Brookman, L., & Koegel, L. K. (2003). Autism: Pivotal response intervention and parental empowerment. *Trends in Evidence-Based Neuropsychiatry, 5(1)*, 61-69.

Krafft, C., Sieverding, M., Salemi, C., & Keo, C. (2018, April). Syrian refugees in Jordan: Demographics, livelihoods, education, and health. In *Economic Research Forum Working Paper Series* (Vol. 1184).

League of Arab States. (2004). *Arab Charter on Human Rights*. Tunis: League of Arab States

Lee, C. (2005). *Empowerment, Access and Rights: Introducing Information and Communication Technology to Women in Costa Rica, Mexico and Nicaragua*. Unpublished PhD Dissertation, The University of Texas at Austin.

Leyland, A. H., & Groenewegen, P. P. (2020). *Multilevel modelling for public health and health services research: health in context* (p. 286). Springer Nature.

Long, S. P., ZHU, X. G., Naidu, S. L., & Ort, D. R. (2006). Can improvement in photosynthesis increase crop yields?. *Plant, cell & environment, 29*(3), 315-330.

Nachshen, J. (2005). Empowerment and families: Building bridges between parents and professionals, theory and research. *Journal on Developmental Disabilities, 1*(11), 67-76.

Nelson-Peterman, J. L., Toof, R., Liang, S. L., & Grigg-Saito, D. C. (2015). Long-term refugee health: health behaviors and outcomes of Cambodian refugee and immigrant women. *Health Education & Behavior*, *42*(6), 814-823.

Norman, P., & Conner, M. (1996). The role of social cognition models in predicting health behaviours: Future directions. In M. Conner & P. Norman (Eds.), *Predicting health behaviour: Research and practice with social cognition models* (pp. 197–225). Open University Press.

Nutbeam, D. (1998). Health promotion glossary. *Health promotion international*, *13*(4), 349-364.

Perkins, D., & Zimmerman, A. (1995). Empowerment Theory, Research and Application. *American Journal of Community Psychology, 23*(5), 569-579.

Pinheiro, I., & Jaff, D. (2018). The role of palliative care in addressing the health needs of Syrian refugees in Jordan. *Medicine, Conflict and Survival*, *34*(1), 19-38.

Price, M., Safadi, R., & Clements, D. (2020). Reproductive health experiences of Syrian refugees residing in Jordan. *Journal of Midwifery and Reproductive Health, 8*(4), 2437-2446.

Razak, A. A., Mansor, N. A., Razak, R. A., Nawi, N. M. M., Yusoff, A. M., & Din, N. (2020). Changing awareness about health behavior: A study among young instagram users. *The Journal of Behavioral Science*, *15*(1), 19-33.

Rechel B, Mladovsky P, Devillé W, Rijks B, Petrova-Benedict R, McKee M. (2011). Migration and health in the European Union. Maidenhead: Open University Press. p. 4. www.euro.who.int/__data/assets/pdf_file/0019/161560/e96458.pdf.

Reynolds, M. (2019). The global refugee "Crisis": learning is the key to solving it. *Soc Educ 83*(6), 308–310.

Riza, E., Kalkman, S., Coritsidis, A., Koubardas, S., Vassiliu, S., Lazarou, D., ... & Linos, A. (2020, June). Community-based healthcare for migrants and refugees: a scoping literature review of best practices. In *Healthcare,* 8(2),115. Multidisciplinary Digital Publishing Institute.

Rizkalla, N., Mallat, N. K., Arafa, R., Adi, S., Soudi, L., & Segal, S. P. (2020). "Children Are Not Children Anymore; They Are a Lost Generation": Adverse Physical and Mental Health Consequences on Syrian Refugee Children. *International journal of environmental research and public health*, 17(22), 8378.

Rousan Tala, Schwabkey Zaker, Jirmanus Lara and Nelson Brett. (2018). Health needs and priorities of Syrian refugees in camps and urban settings in Jordan: Perspectives of refugees and health care providers. *Eastern Mediterranean Health Journal, 24* (3), 243- 253.

Samara, A., Sweis, R. J., Tarawneh, B., Albalkhy, W., Sweis, G., & Alhomsi, S. (2022). Sustainability management of international development projects by International Non-Governmental Organizations: The case of INGOs working with refugees in Jordan. *International Journal of Construction Management*, 22(9), 1657-1666.

Scholz, N. (2016). The public health dimension of the European migrant crisis. *EPRS/ European Parliamentary Research Service Brief*, 8.

Serrani Azcurra D.J.L. (2014) Elders Health Empowerment Scale. Spanish adaptation and psychometric analysis. *Colomb Med, 45*(4), 179-85.

Shah, S., Siddharth, M., & Yun, K. (2020). Primary Care of Refugee Children. In *Refugee Health Care* (285-302). Springer, Cham.

Sharp, M., Parpia, A., Ahram, M., Mahmoud, R., & Khoshnood, K. (2021). Prevalence of and risk factors for depression among female Syrian refugees and Jordanians with chronic disease: a pilot study. *Eastern Mediterranean Health Journal*, 27(12), 1129 - 1151.

Shatnawi, A.M., AlSobeh, A.M., AL-Mifleh, E.I. and Migdady, A.F., 2022. The Effectiveness of a Program Based on Psychosocial Support in Raising the Level of Family Empowerment among Refugees in Jordan. *International Journal of Psychological and Educational Research*, 1(4).

Shearer, N., & Fleury, J. (2006). Social support promoting health in older women. *Journal of Women & Aging*, *18*(4), 3-17.

Simpson, V. (2015). Models and theories to support health behavior intervention and program planning. *Health and Human Sciences*, *2*(8), 45-47.

Sørensen, K., Van den Broucke, S., Fullam, J., Doyle, G., Pelikan, J., Slonska, Z., & Brand, H. (2012). Health literacy and public health: a systematic review and integration of definitions and models. BMC public health, 12(1), 1-13.

Steimel, S. (2017). Negotiating refugee empowerment (s) in resettlement organizations. *Journal of Immigrant & Refugee Studies, 15*(1), 90-107.

Stevens, D. (2016). Rights, needs or assistance? The role of the UNHCR in refugee protection in the Middle East. *The International Journal of Human Rights*, *20*(2), 264-283.

Strong, J., Varady, C., Chahda, N., Doocy, S., & Burnham, G. (2015). Health status and health needs of older refugees from Syria in Lebanon. *Conflict and health*, *9*(1), 1-10.

Sudikno, S., Sulistyowati, N., Nainggolan, O., & Tjandrarini, D. H. (2021). Obesity in Older Adults Indonesia: The Role of Healthy Behaviour Factors and Metabolic Syndrome. *Global Journal of Health Science*, *13*(6), 124-124.

The European Council on Refugees & Exiles. (2002). *Position on the integration of refugees in Europe*. Retrieved from http://www.ecre.org/component/downloads/downloads/168.html

Thoresen, P., Fielding, A., Gillieatt, S., Thoresen, S. (2016). Identifying the needs of refugee and asylum-seeking children in Thailand: a focus on the perspectives of children. *J Refugee Stud 30*(3), 01-28.

Tiltnes, Å. A., Zhang, H., & Pedersen, J. (2019). The living conditions of Syrian refugees in Jordan. *FAFO Report. Amman: Fafo Research Foundation.*

Tomlinson, F., & Egan, S. (2002). From Marginalization to (Dis)Empowerment: Organizing Training and Employment Services for Refugees. *Human Relations, 55*(8), 1019-1043.

UNHCR. (2018). *Global trends: Forced displacement in 2017*. Geneva: UNHCR.

Verme, P., Gigliarano, C., Wieser, C., Hedlund, K., Petzoldt, M., & Santacroce, M. (2015). *The welfare of Syrian refugees: evidence from Jordan and Lebanon*. World Bank Publications.

Webb, T. L., Sniehotta, F. F., & Michie, S. (2010). Using theories of behaviour change to inform interventions for addictive behaviours. *Addiction*, 105(11), 1879-1892.

Wells, R.; Steel, Z.; Abo-Hilal, M.; Hassan, A. H. & Lawsin, C. (2016). Psychosocial concerns reported by Syrian refugees living in Jordan: Systematic review of unpublished needs assessments. *The British Journal of Psychiatry,* 209 (2), 99–106. doi:10.1192/bjp.bp.115.165084

World Health Organization. (2016). *Refugees and migrants most common health problems, migration and health key issues.* World Health Organization. Retrieved from http://www.euro.who.int/en/health-topics/health-determinants/migration-and-health/migrant-health-in-the-european-region/migration-and-health-key-issues.

Yang, Y. M., Wang, H. H., Lee, F. H., Lin, M. L., & Lin, P. C. (2015). Health empowerment among immigrant women in transnational marriages in Taiwan. *Journal of Nursing Scholarship*, *47*(2), 135-142.

Zimmerman, A. (1995). Psychological empowerment: Issues and illustrations. *American Journal of Community Psychology, 23*(5), 581–600.

Zimmerman, A. (2000). *Empowerment theory: Psychological, organizational, and community levels of analysis*. In J. Rappaport & E. Seidman (Eds.), Handbook of community psychology (pp. 43–63). New York: Kluwer Academic/Plenum.

Zuckerman, M., Knee, C., Kieffer, S., & Gagne, M. (2004). What individual believe they can or cannot do? Exploration of realistic and unrealistic control beliefs. *Journal of Personality Assessment, 82*(2), 215–232.

Zumbo, B., Kelly, K., Begoray, D., Kazanjian, A., Mullet, J., & Hayes, M. (2006). The development and validation of measures of "Health Literacy" in different populations. *UBC Institute of Health Promotion Research and UVic Community Health Promotion Research: University of British Columbia*.

# فاعلية برنامج تدريبي مستند إلى التثقيف الصحي في رفع مستوى التمكين الصحي لدى اللاجئين في الأردن

احمد الشريفين، قسم علم النفس الإرشادي والتربوي، كلية التربية، جامعة اليرموك، الأردن

ahmed.sh@yu.edu.jo

Orcid: 0000-0003-0617-5998

مؤيد مقدادي، قسم علم النفس الإرشادي والتربوي، كلية التربية، جامعة اليرموك، الأردن

muayyad@yu.edu.jo

ORCID: 0000-0003-0767-6153

أماني شطناوي، قسم تكنولوجيا المعلومات، كلية تكنولوجيا معلومات وعلوم الحاسوب، جامعة اليرموك، الأردن

ashatnawi@yu.edu.jo

orcid: 0000-0002-5096-9405

أنس "محمد رمضان" الصبح (المؤلف المراسل)

قسم نظم المعلومات، كلية تكنولوجيا المعلومات وعلوم الحاسوب، جامعة اليرموك، الأردن anas.alsobehs@yu.edu.jo

قسم تكنولوجيا المعلومات، كلية الحاسوب، جامعة جنوب ولاية الينوي، الولايات المتحدة الامريكية، anas.alsobseh@siu.edu

orcid: 0000-0002-1506-7924

أية عكاوي، قسم اللغة الإنجليزية وادابها، كلية الآداب، جامعة اليرموك، الأردن

aya-aka @yu.edu.jo

**الملخص:**

**الأهداف:** هدفت الدراسة إلى تقييم فعالية برنامج التدريب القائم على التثقيف الصحي في تعزيز مستوى التمكين الصحي بين اللاجئين في الأردن. يعد التمكين الصحي مكوناً رئيساً لتعزيز الصحة، حيث إنه يمكّن الأفراد من التحكم في صحتهم وإدارة نتائجهم الصحية وتحسينها. اللاجئون هم فئة من الفئات السكانية المستضعفة والتي لها محدودية الوصول إلى الرعاية الصحية.

**المنهجية:** تكونت عينة الدراسة من 38 لاجئاً في محافظة إربد، الأردن، والذين تم اختيارهم بالطريقة المتيسرة من خلال التنسيق مع بعض المنظمات العاملة في مجال اللجوء في المحافظة، إذ تقسيمهم عشوائياً إلى مجموعتين: مجموعة تجريبية (ن = 19) تلقت برنامج التدريب على التثقيف الصحي، ومجموعة ضابطة (ن = 19) لم تتلقَّ أي تدريب على التثقيف الصحي. استُخدم مقياس التمكين الصحي (HES)، وهو أداة تم التحقق من صحتها، بغرض جمع البيانات من كلا المجموعتين في الاختبارات القبلية والبعدية، وتم إجراء اختبار متابعة لأعضاء المجموعة التجريبية فقط.

**النتائج:** أظهرت النتائج زيادة ذات دلالة إحصائية في درجات التمكين الصحي للمجموعة التجريبية التي تلقت البرنامج التدريبي مقارنة بالمجموعة الضابطة؛ حيث كان الوسط الحسابي للاختبار القبلي للمجموعة التجريبية (1.97 ± 0.27)، وللمجموعة الضابطة (1.84 ± 0.21)، وأصبح للاختبار البعدي في المجموعة التجريبية (3.88 ± 0.13)، بينما للمجموعة الضابطة (1.85 ± 0.20). كما أظهر اختبار المتابعة الحفاظ على مستويات التمكين الصحي المعززة في المجموعة التجريبية، مع عدم وجود فرق كبير بين درجات الاختبار البعدي والمتابعة؛ مما يدل على فاعلية برنامج التدريب على التثقيف الصحي في تعزيز التمكين الصحي للاجئين في الأردن.

**الخلاصة:** هناك مسؤوليات كبيرة على مقدمي الرعاية الصحية وصانعي السياسات في الأردن وغيرها من الدول التي تضم أعداداً هائلة من اللاجئين، ومن الضروري إعطاء أولوية كافية للتعليم الصحي وبرامج النشاط التوعوي لتعزيز التمكين الصحي للاجئين، وتحسين ثقافتهم الصحية؛ مما ينعكس على الصحة العامة لديهم.

**الكلمات الدالة:** اللاجئون، التثقيف الصحي، التمكين الصحي، السكان المستضعفون، الأردن.